\documentclass{PoS}
\usepackage[utf8]{inputenc}
\usepackage{graphicx}
\usepackage{dcolumn}
\usepackage{bm}
\usepackage{amsmath}
\usepackage{amssymb}

\def\be{\begin{equation}}
 \def\ee{\end{equation}}
  \def\bea{\begin{eqnarray}}
 \def\eea{\end{eqnarray}}

\newcommand{\eq}[1]{(\ref{#1})}
\newcommand{\jpsi}{J/\psi}
\newcommand{\dd}{{\rm d}}
\newcommand{\sqrts}{\sqrt{s}}
\newcommand{\pt}{p_{_\perp}}
\newcommand{\ptmin}{p_{_\perp}^\text{min}}
\newcommand{\dsigpp}{\dd\sigma_{\rm pp}}

\newcommand{\RAA}{\ensuremath{R_{_{\rm AA}}}}

\newcommand{\qhat}{\hat{q}}

\newcommand\omc{\omega_c}
\newcommand\omcp{{{\bar{\omega}}_c}}
\newcommand\meanz{\langle z \rangle}
\newcommand\meaneps{\langle \epsilon \rangle}
\newcommand\meanepsbar{\langle \bar{\epsilon} \rangle}

\title{Quenching of hadron spectra in XeXe and PbPb collisions at the LHC}

\ShortTitle{Quenching of hadron spectra in XeXe and PbPb collisions at the LHC}

\author{\speaker{Fran\c{c}ois Arleo}, Guillaume Falmagne\\
        Laboratoire Leprince-Ringuet, \'Ecole polytechnique, CNRS/IN2P3,  Universit\'e Paris-Saclay,  91128, Palaiseau, France\\
        E-mail: \email{francois.arleo@cern.ch}, \email{guillaume.falmagne@cern.ch}}

\abstract{The $\pt$ dependence of the nuclear modification factor $\RAA$ measured in PbPb collisions at the LHC exhibits a universal shape, which can be very well reproduced in a simple energy loss model based on the BDMPS medium-induced gluon spectrum. We update a former study by including in the analysis the recent CMS measurements on $\jpsi$ production in PbPb collisions at $\sqrts=5.02$~TeV and charged hadron production in XeXe collisions at $\sqrts=5.44$~TeV. The average parton energy loss extracted from minimum bias XeXe collisions is reduced typically by 20\% compared to PbPb collisions, consistent with a length dependence $\langle\bar{\epsilon}\rangle \propto L^{1.3\pm0.5}$ with $L\propto A^{1/3}$.}

\FullConference{International Conference on Hard and Electromagnetic Probes of High-Energy Nuclear Collisions\\
		30 September - 5 October 2018\\
		Aix-Les-Bains, Savoie, France}

\begin{document}

\section{Introduction}

The depletion of high-$\pt$ hadron spectra measured in heavy ion collisions with respect to pp collisions has provided strong evidence of radiative energy loss of quarks and gluons propagating in quark-gluon plasma (QGP). The pioneering measurements at RHIC followed by those at LHC have allowed for a rich phenomenology based on a variety of theoretical frameworks and underlying assumptions (see Ref.~\cite{Qin:2015srf} for a recent review). However, computing the $\pt$ dependence of the nuclear modification factor $\RAA$ presents a formidable challenge as many physical processes come into play.
Recently, we have instead attempted to describe the hadron quenching at large $\pt$ (typically, $\pt\gtrsim 10$~GeV), assuming \emph{only} radiative energy loss to be at work~\cite{Arleo:2017ntr}. It was demonstrated that the $\pt$ dependence of $\RAA$ predicted in a simple analytic model, based on the BDMPS or GLV energy loss formalism, proves in excellent agreement with charged hadron data in PbPb collisions at $\sqrts=2.76$~TeV and $\sqrts=5.02$~TeV and in all centrality classes.\footnote{Except the most peripheral data sets, possibly affected by event selection and geometry biases~\cite{Morsch:2017brb}. It would be interesting to analyze as well these data sets taking into account this bias.} The universality of charged hadron quenching observed in Ref.~\cite{Arleo:2017ntr} for various centrality classes and at different energies supports this picture. In addition, it was shown that the quenching of  $J/\psi$ and D mesons follows the same trend as that of charged hadrons, suggesting a possible common origin for the depletion of \emph{all} hadron species at large transverse momentum. 

Naturally, more precise measurements on a wider range in transverse momentum, for several particle species and in different collision systems would help to further investigate the scaling properties of hadron quenching in heavy ion collisions. In that spirit, we analyze in these proceedings the quenching of $\jpsi$ production in PbPb collisions at $\sqrt{s}=5.02$~TeV~\cite{Sirunyan:2017isk} and that of charged hadrons in XeXe collisions at $\sqrt{s}=5.44$~TeV~\cite{Sirunyan:2018eqi}, recently made available by the CMS collaboration and shown at this conference. 

\section{Model and method}

According to the energy loss model, the nuclear modification factor at large $\pt$ reads~\cite{Arleo:2017ntr}
\be\label{eq:RAA2}
\RAA^{h}(\pt) = \int_0^{\infty}\,\dd\epsilon\  \left( 1 + \frac{\epsilon}{\pt} \right)^{-n}\ {P}(\epsilon) = \int_0^{\infty}\,\dd{x} \  \left( 1 + \frac{x \omcp}{\pt} \right)^{-n}\ \bar{P}(x) 
\ee
where ${P}$ represents the probability density for the propagating particle to lose the energy $\epsilon$ while traversing the hot medium; it is a scaling function of the energy loss scale $\omcp$, $P(\epsilon) \equiv 1/\omcp \,\bar{P}(\epsilon/\omcp)$. In \eq{eq:RAA2}, $n$ is the power law index of the hadron spectrum in pp collisions, $\dsigpp^{h}/\dd\pt\propto\pt^{-n}$. Performing the replacement $(1+x\omcp/\pt)^{-n}$ by $\exp(-n\,x\omcp/\pt)$, as suggested by Baier {\it et al.}~\cite{Baier:2001yt}, leads to an approximate scaling in the variable $\pt/\,n\omcp$, which now allows for comparing the values of $\RAA$ at different center-of-mass energies and for different hadron species.
In the present analysis, $\RAA(\pt)$ is computed numerically from \eq{eq:RAA2} using the quenching weight determined in~\cite{Arleo:2002kh} from the BDMPS medium-induced gluon spectrum~\cite{Baier:2001yt}.

In this simple energy loss model, the {\it shape} of $\RAA$ as a function of $\pt$ is thus fully predicted once the exponent $n$ is known, obtained from a fit to pp data at the corresponding center-of-mass energy ($n$ ranges from $n=5.2$ to $n=5.9$ depending on the particle species and collision energies). What remains to be determined is the energy loss scale $\omcp$, or equivalently the first moment of the quenching weight, which depends in principle on the space-time evolution of the QGP energy density and the geometry of the heavy ion collision. Rather than modeling the hot medium, the value of $\omcp$ is obtained from 1-parameter fits to each data set, in a given centrality class and at a given $\sqrts$, from $\ptmin=10$~GeV onwards (in the most central sets, where $\omcp$ is large, we take $\ptmin=15$~GeV). Data include charged hadron production measured by CMS in PbPb collisions in five centrality classes at both colliding energies~\cite{CMS:2012aa,Khachatryan:2016odn}. In this update, we have also analyzed the CMS measurements of $\jpsi$ production in PbPb collisions at $\sqrts=5.02$~TeV~\cite{Sirunyan:2017isk} and charged hadron production in XeXe collisions at $\sqrts=5.44$~TeV~\cite{Sirunyan:2018eqi}.

\section{Results and discussion}

%%%%%%%%%%%%%%%%%%%%%%%%%%%%%%%%%%%%%%%%%%
\begin{figure}[tbp]
\begin{center}
    \includegraphics[width=7.3cm]{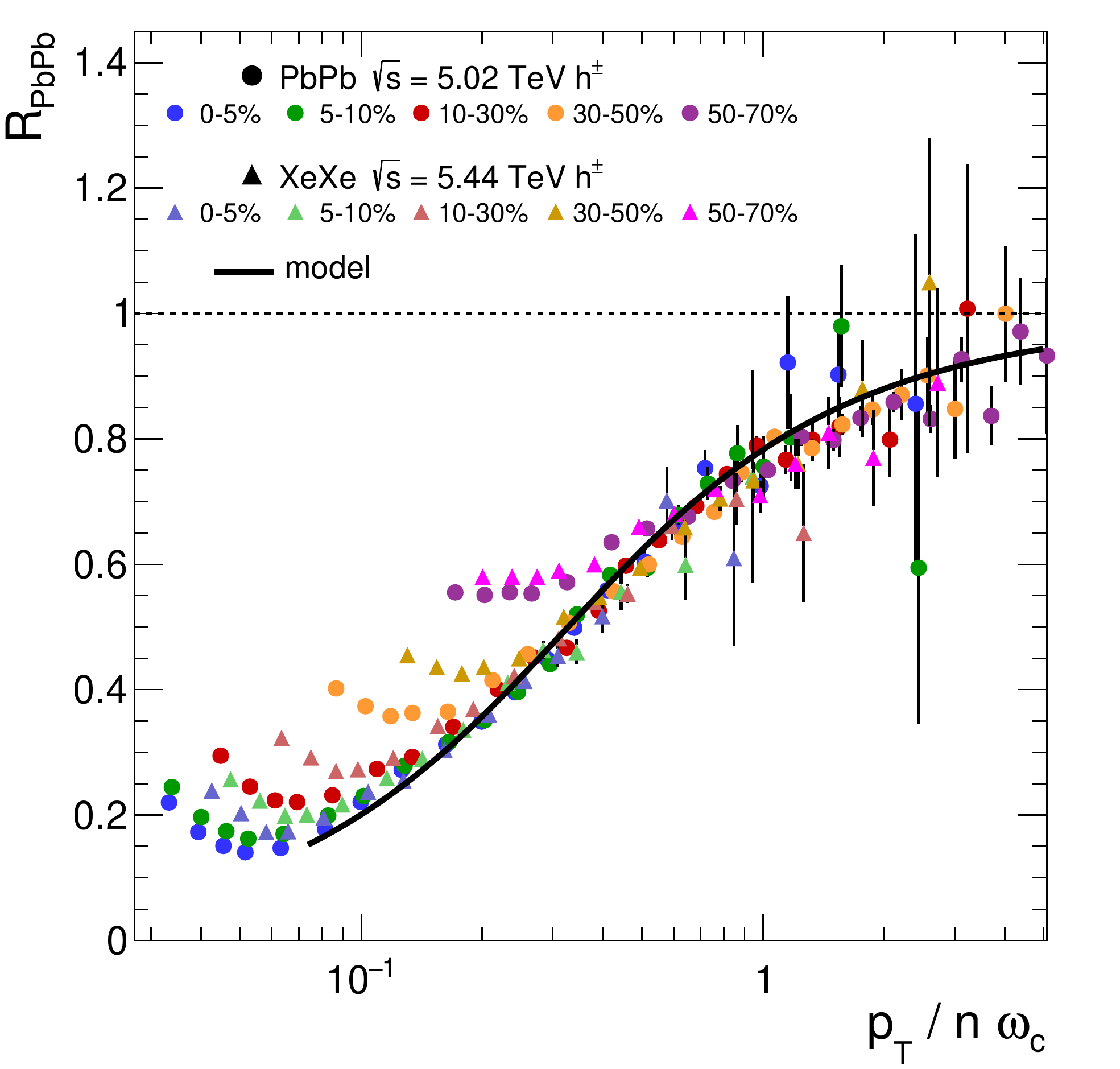}
    \includegraphics[width=7.3cm]{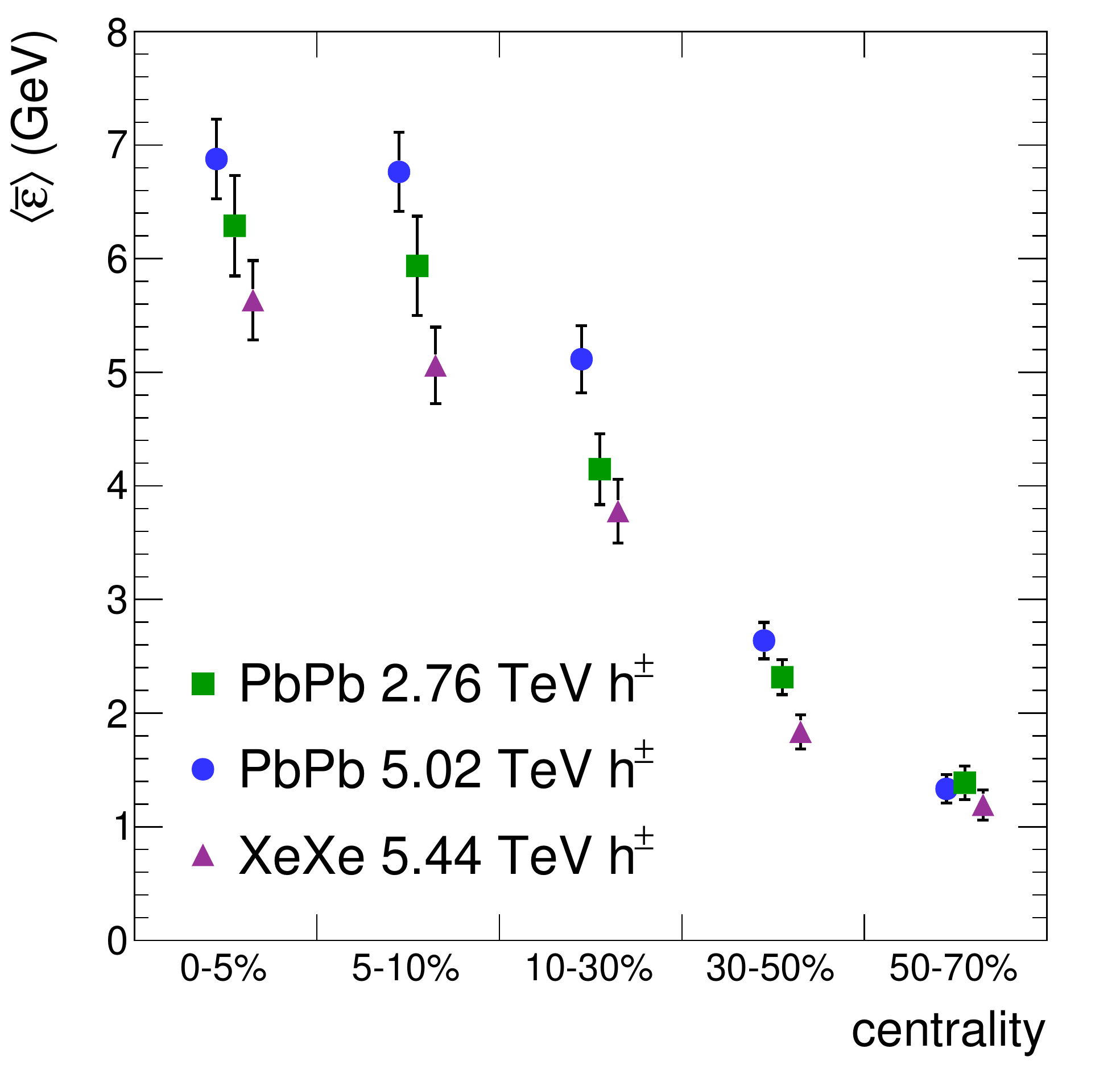}
  \end{center}
\vspace{-0.4cm}
\caption{Left: $\RAA$ of charged hadrons as a function of $\pt/\,n\omc$ in PbPb ($\sqrts=5.02$~TeV) and in XeXe ($\sqrts=5.44$~TeV) collisions in different centrality classes (for clarity only statistical uncertainties are shown). Right: Mean energy loss extracted from charged hadron data in PbPb and XeXe collisions.}
   \label{fig:scaling}
\end{figure}
%%%%%%%%%%%%%%%%%%%%%%%%%%%%%%%%%%%%%%%%%%

Let us first discuss the new results coming from the analysis of charged hadron quenching in XeXe collisions at $\sqrt{s}=5.44$~TeV. The shape of $\RAA(\pt)$ given by the model proved in good agreement with CMS measurements in all centrality classes (as in Ref.~\cite{Arleo:2017ntr} we exclude the most peripheral data set).
In order to exhibit the universality of hadron quenching in different collision systems, Fig.~\ref{fig:scaling} (left) shows all charged hadron data points in PbPb and XeXe (respectively at $\sqrt{s}=5.02$ and $5.44$~TeV) plotted as a function of the scaling variable, $\pt/\,n\omcp$, together with the shape of $\RAA$ from Eq.~\eq{eq:RAA2}.
All data exhibit the predicted scaling, supporting the interpretation of a unique process responsible for the quenching of charged hadrons above a given $\pt$ in different collision systems (PbPb, XeXe) and various centrality classes.
Interestingly, scaling violations can be seen at low $\pt/n\omcp$ for all centralities, corresponding to particle production with transverse momenta $\pt \lesssim 10$~GeV. This may signal the onset of other phenomena below this scale.

Apart from investigating the scaling of $\RAA$ for different hadron species or collision systems, this procedure allows for extracting the average energy loss (times the mean fragmentation variable), $\meanepsbar \equiv \meanz\times\meaneps$, experienced by the fast parton in the QGP, as a function of centrality. 
As can be seen in Fig.~\ref{fig:scaling} (right), $\langle \bar{\epsilon} \rangle$ is (as expected) maximal in the most central bins of PbPb collisions at $\sqrt   {s}=5.02$~TeV, $\langle \bar{\epsilon}\rangle \simeq6.9$~GeV. It starts to drop in the more peripheral classes, reaching $\langle \bar{\epsilon}\rangle \simeq1$~GeV in the $50$--$70\%$ centrality class. It is particularly interesting to note that the mean energy loss in PbPb collisions at $\sqrts=5.02$~TeV is roughly $10$--$20\%$ larger than at $\sqrts=2.76$~TeV. This is nicely consistent with the measurements of ALICE~\cite{Adam:2015ptt} which show that the multiplicity density, $dN/dy \propto \qhat \propto \meaneps$, increase by roughly the same amount.
It is also instructive to compare the energy loss scale obtained in PbPb and XeXe collisions at two nearby collision energies. Without much surprise, the mean energy loss values extracted from XeXe data are systematically lower (by roughly 20\%) than those obtained from PbPb measurements, dropping from $\langle \bar{\epsilon} \rangle\simeq5.6$~GeV in most central collisions down to $\langle \bar{\epsilon} \rangle\simeq1$~GeV in the centrality class $50$-$70\%$.
In addition, we have analyzed the centrality-integrated charged hadron data in both PbPb and XeXe collisions systems (class $0$-$100\%$ and $0$-$80\%$, respectively). Results are ${\langle \bar{\epsilon} \rangle}_{\text{PbPb}}=4.5\pm0.2$~GeV and ${\langle \bar{\epsilon} \rangle}_{\text{XeXe}}=3.7\pm0.2$~GeV. Assuming that the average energy loss scales like a power of the (naive) medium length scale $L\sim A^{1/3}$, we infer empirically from these results that ${\langle \bar{\epsilon} \rangle} \propto A^{\beta/3}$ with $\beta=1.3\pm0.5$. The value of this exponent could reflect the pathlength dependence of parton energy loss as well as the difference of transport coefficient from one system to another.

%%%%%%%%%%%%%%%%%%%%%%%%%%%%%%%%%%%%%%%%%%
\begin{figure}[tbp]
\begin{center}
    \includegraphics[width=7.3cm]{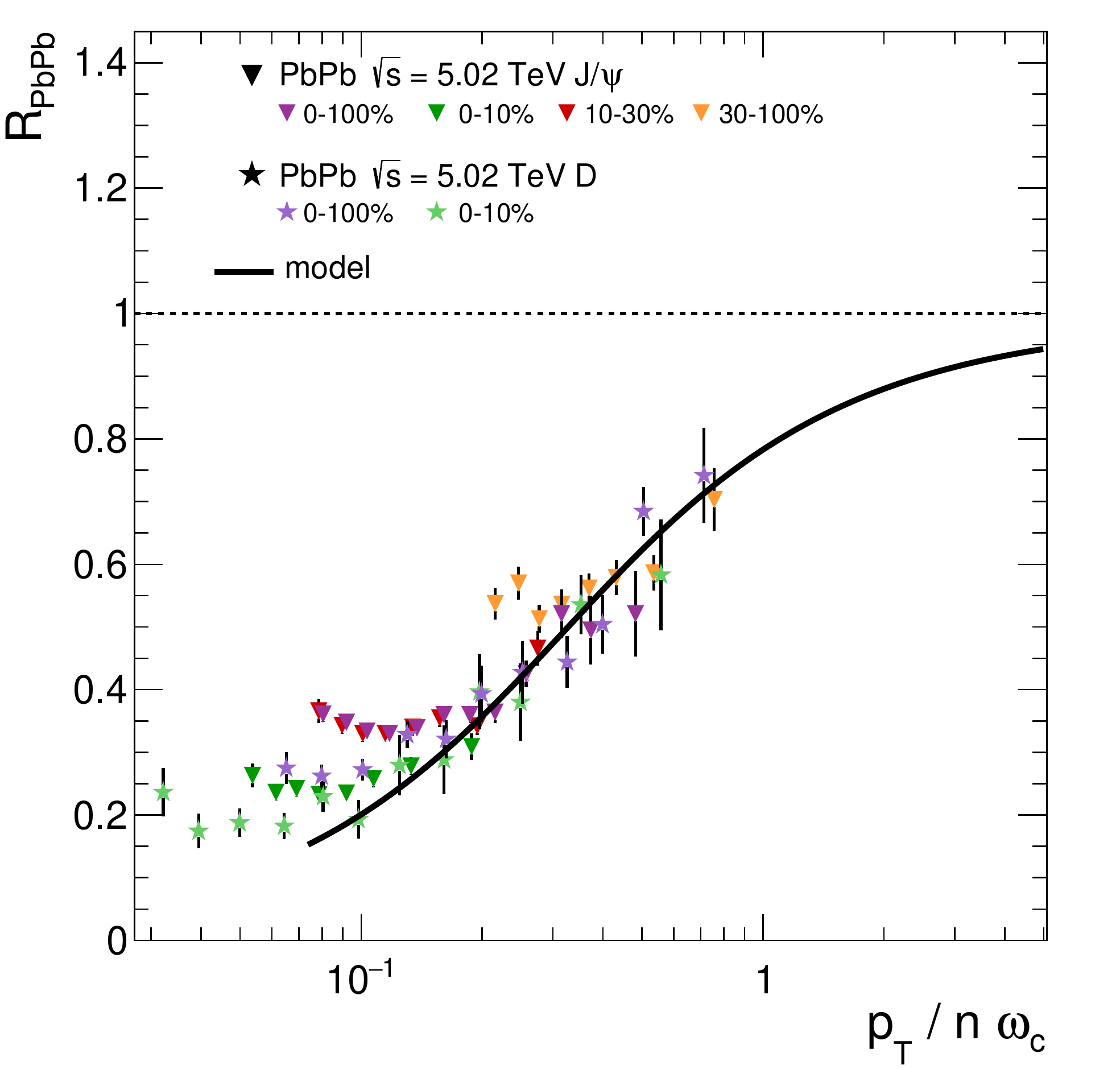}
    \includegraphics[width=7.3cm]{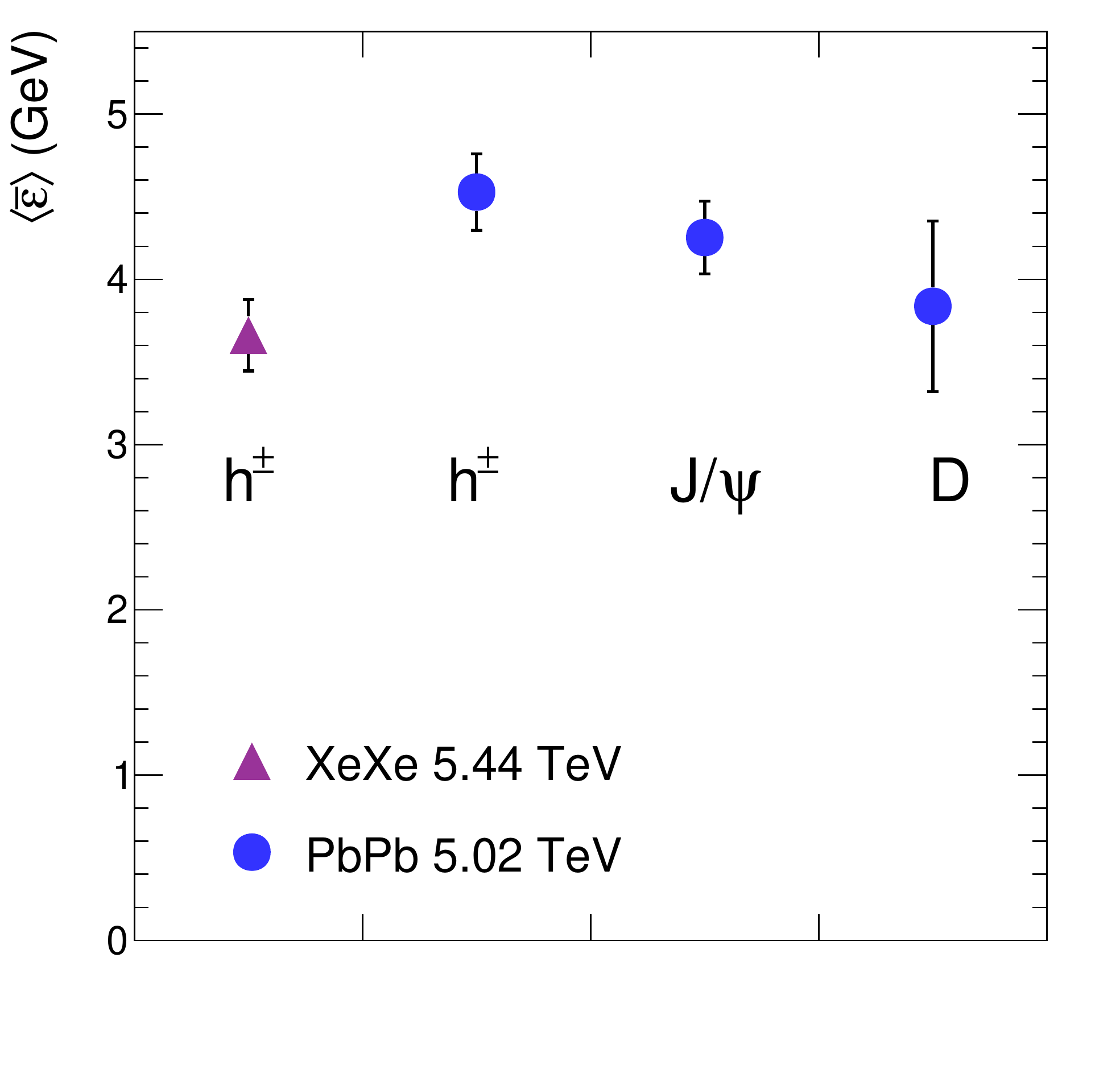}
  \end{center}
\vspace{-0.4cm}
\caption{Left: $\RAA$ of D and $\jpsi$ mesons as a function of $\pt/\,n\omc$ in PbPb collisions at $\sqrts=5.02$~TeV and in different centrality classes (for clarity only statistical uncertainties are shown). Right: Mean energy loss extracted from minimum bias PbPb and XeXe collisions for different particle species.}
   \label{fig:scaling2}
\end{figure}
%%%%%%%%%%%%%%%%%%%%%%%%%%%%%%%%%%%%%%%%%%

Moving to the heavy quark sector, the recently available CMS measurements of D~\cite{Sirunyan:2017xss} and $\jpsi$~\cite{Sirunyan:2017isk} mesons in PbPb collisions at $\sqrt{s}=5.02$~TeV have also been fitted within the same energy loss model.\footnote{Results on $\jpsi$ from ATLAS~\cite{Aaboud:2018quy} and coming D meson data from ALICE should be included in a future analysis.}  Fig.~\ref{fig:scaling2} (left) shows these data as a function of $\pt/n\omega_c$. 
A good agreement is observed within experimental uncertainties, for almost all centrality classes, except perhaps the $\jpsi$ most central $0$-$10\%$ data set for which the $\pt$ dependence is more pronounced in the model than in data. More generally, more precise data at at higher $\pt$ would be required to check further the model and the similarity between light and heavy hadrons. These results nevertheless suggest that at large $\pt$ the same process --~namely radiative energy loss~-- affects similarly all hadron species, including bound states like heavy-quarkonia~\cite{Arleo:2017ntr}.

The values of $\meanepsbar$ extracted from the quenching of charged hadrons, $D$ and $\jpsi$ production in centrality integrated PbPb collisions at $\sqrt{s}=5.02$~TeV and charged hadron production in (0-80\%) centrality integrated XeXe collisions at $\sqrt{s}=5.44$~TeV are shown in Fig.~\ref{fig:scaling2} (right). Perhaps a bit surprisingly at first glance, no genuine difference between the energy loss scale in the D and $\jpsi$ channels is observed. Assuming that $D$ and $\jpsi$ production at large $\pt$ come from charm quark and gluon fragmentation, respectively, a naive estimate  would lead to ${\meaneps}_{\jpsi} / {\meaneps}_{D} = C_A/C_F=9/4$. No such factor would be expected, of course, if gluon fragmentation processes dominate the production of D mesons.
Compared to charged hadrons, the average energy loss from $\jpsi$ data is similar within uncertainties (${\langle \bar{\epsilon} \rangle}^{J/\psi}_{\text{PbPb}}=4.3\pm0.2$~GeV vs. ${\langle \bar{\epsilon} \rangle}^{h^\pm}_{\text{PbPb}}=4.5\pm0.2$~GeV). The energy loss in the D meson channel (${\langle \bar{\epsilon} \rangle}^{D}_{\text{PbPb}}=4.1\pm0.7$~GeV) is also comparable in magnitude.

\providecommand{\href}[2]{#2}\begingroup\raggedright\endgroup

\end{document}